# Formally Verifying WARP-V, an Open-Source TL-Verilog RISC-V Core Generator

Steven Hoover
Redwood EDA
Shrewsbury, MA, USA
(774)773-8333
steve.hoover@redwoodeda.com

Ákos Hadnagy
Delft, The Netherlands
akos.hadnagy@gmail.com

**Biographies**

Steve Hoover is the founder of Redwood EDA, provider of Makerchip.com. He is an 18-year veteran chip designer, having led contributions to DEC Alpha, and Intel Itanium, x86, and OmniPath network chips. Steve is an active driver of the TL-Verilog language definition and is active in numerous TL-Verilog-related open-source projects.

Ákos Hadnagy is a master's student at TU Delft. He became involved in the WARP-V project through the Google Summer of Code program this summer. His interests include heterogeneous and reconfigurable computing, FPGAs and hardware development.

**Abstract**

*Abstract*— **Timing-abstract and transaction-level design using TL-Verilog have shown significant productivity gains for logic design. In this work, we explored the natural extension of transaction-level design methodology into formal verification.**

**WARP-V is a CPU core generator written in TL-Verilog. Our primary verification vehicle for WARP-V was a formal verification framework for RISC-V, called riscv-formal. The timing-abstract and transaction-level logic modeling techniques of TL-Verilog greatly simplified the task of creating a harness connecting the WARP-V model to the verification interface of riscv-formal. Furthermore, the same harness works across all RISC-V configurations of WARP-V.**

## I. Introduction

System-on-chip (SoC) design methodology employs principles of modularity and reuse to managing the complexity of modern silicon. For SoC building blocks, known as intellectual property (IP), to be broadly reusable, they must be applicable under a range of design constraints. Register-transfer level (RTL) methodology, which predates SoC design, is ill-suited to delivering this flexibility. Transaction-Level Verilog (TL-Verilog) has been shown to provide flexibility through a timing-abstract modeling approach [1]. At the same time, unlike high-level synthesis, TL-Verilog allows designers to retain direct RT-Level control.

WARP-V [2] is a RISC-V core written in TL-Verilog that achieves an unprecedented level of architectural flexibility and scalability in a small amount of code. It can implement a single-stage CPU, appropriate as a microcontroller, or a seven-stage mid-range general-purpose processor. To fully deliver flexible IP, flexible verification modeling is also required. This paper presents the first real-world use of TL-Verilog for verification modeling and shows how the flexibility benefits of TL-Verilog modeling extend into verification.

The remainder of this paper is organized as follows. The WARP-V model and its verification harness are described in Section II. Section III quantifies the benefits, and Section IV summarizes the contributions.

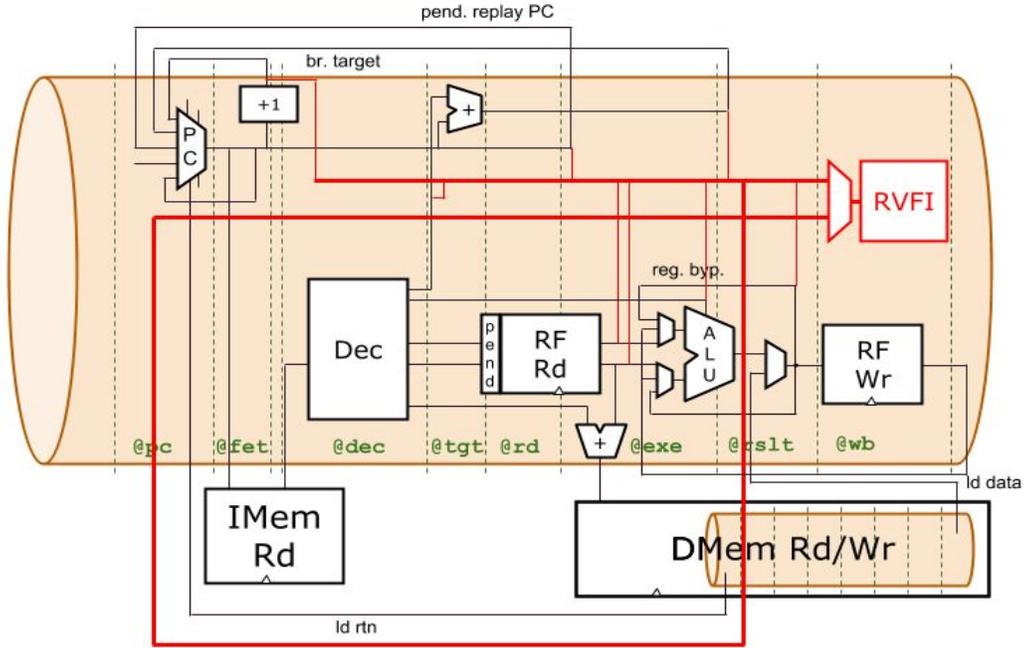

Fig. 1. WARP-V microarchitecture with verification harness

## I. WARP-V and its Verification Harness

WARP-V employs a traditional scalar (currently), pipelined microarchitecture, depicted as the black logic in Fig. 1. WARP-V provides pipeline flexibility through the use of "virtual" pipeline stages, illustrated in green. These virtual stages can all be mapped to a single physical pipeline stage to provide a single-cycle implementation, or they can be mapped to different physical stages for a more deeply pipelined implementation.

WARP-V was brought to life using a single 11-instruction test program. The remaining verification was entirely done through formal verification using a framework called riscv-formal [3]. Riscv-formal is an open-source formal verification framework for RISC-V cores. It provides assertions written in Verilog to be verified by the SymbiYosys model checker [4]. Designs interface with riscv-formal using the RISC-V Formal Interface (RVFI). This interface takes in instructions, which can be presented out-of-order, tagged with an instruction-order identifier. Each instruction is accompanied by a number of fields characterizing the instruction and its side effects.

Red logic in Fig. 1 shows the RVFI and the test harness necessary to connect signals of WARP-V to the RVFI. Signals related to a given instruction must be presented to the RVFI in the same cycle, but the signals of an instruction in the WARP-V design that are used to provide these inputs are distributed across different stages of the pipeline. This can be seen in Fig. 1 as the red wires feeding into the bus that feeds the upper RVFI multiplexer input. A single-stage WARP-V requires no signal staging (flip-flops) for this, but a deeply pipelined WARP-V requires the correct number of flip-flops to align each input. This is exactly the need that is addressed by TL-Verilog's timing-abstract modeling [1]. RVFI signals can be driven by logic expressions in pipeline stage @wb, and these will imply correct staging of input signals. Pipeline flexibility requires no special consideration when connecting the RVFI inputs as proper staging is implemented by tools.

It is not, however, always the instruction in the @wb stage that must be provided to the RVFI. The WARP-V microarchitecture supports a memory with arbitrary access latency. When a load instruction reads data from memory, a pseudo-load-return instruction is injected into the pipeline to reserve a slot to write the load data into the register file. Loads must be presented to the RVFI relative to the pseudo-load-return instruction as the load data may not be available in time to present the original load instruction. The remaining fields presented to the RVFI, however, must come from the original load instruction. They must, therefore, be recirculated back to the pseudo-load-return instruction to again flow through the pipeline and be presented to the RVFI. This is depicted in Fig. 1 as the lower red bus that recirculates through the pipeline into the lower RVFI multiplexer input.

```
@M4_NEXT_PC_STAGE
   ?$returning_ld
      /original_ld
         $ANY = /top|mem/data>>M4_LD_RETURN_ALIGN$ANY;
@M4_REG_WR_STAGE
   /original
      $ANY = /instr$returning_ld ?
               /instr/original_ld$ANY :
               /instr$ANY;
```

Fig. 2. Load instruction recirculation and mux code

```
/original
   *rvfi_insn      = $raw;
   *rvfi_order     = $rvfi_order;
   *rvfi_rs1_addr  = /src[1]$is_reg ? $raw_rs1 : 5'b0;
   *rvfi_rs2_addr  = /src[2]$is_reg ? $raw_rs2 : 5'b0;
```

Fig. 3. RVFI signal assignments

This need is addressed by another important aspect of TL-Verilog--transaction flow. Fig. 2 shows the code defining this recirculation flow and the RVFI multiplexer. (These are not together in the real code, as the recirculation code is part of the design, and the multiplexer is part of the verification harness.) The multiplexer normally selects the instruction in the pipeline, but, for pseudo-load-return, it selects the original load instruction through the recirculation flow. Though the exact mechanism and syntax are outside of the scope of this paper, details can be found in the TL-X 2a HDL Extension Draft Syntax Specification [5]. The important takeaway is that the recirculation and selection of any needed signals is automated with minimal code. The instruction we present to the RVFI is in the /original scope. Fig. 2 shows assignments in this scope to SystemVerilog RVFI signals which naturally consume signals associated with the correct instruction. (Note that many RVFI inputs are not relevant to loads. We do avoid unnecessary recirculation of these signals by connecting them directly from the instruction in the pipeline without going through the multiplexer.)

## I. Results

The entire test harness is implemented in 59 lines of commented code. It connects the WARP-V model with 19 RVFI signals. The red wiring in Fig. 1 is automated.

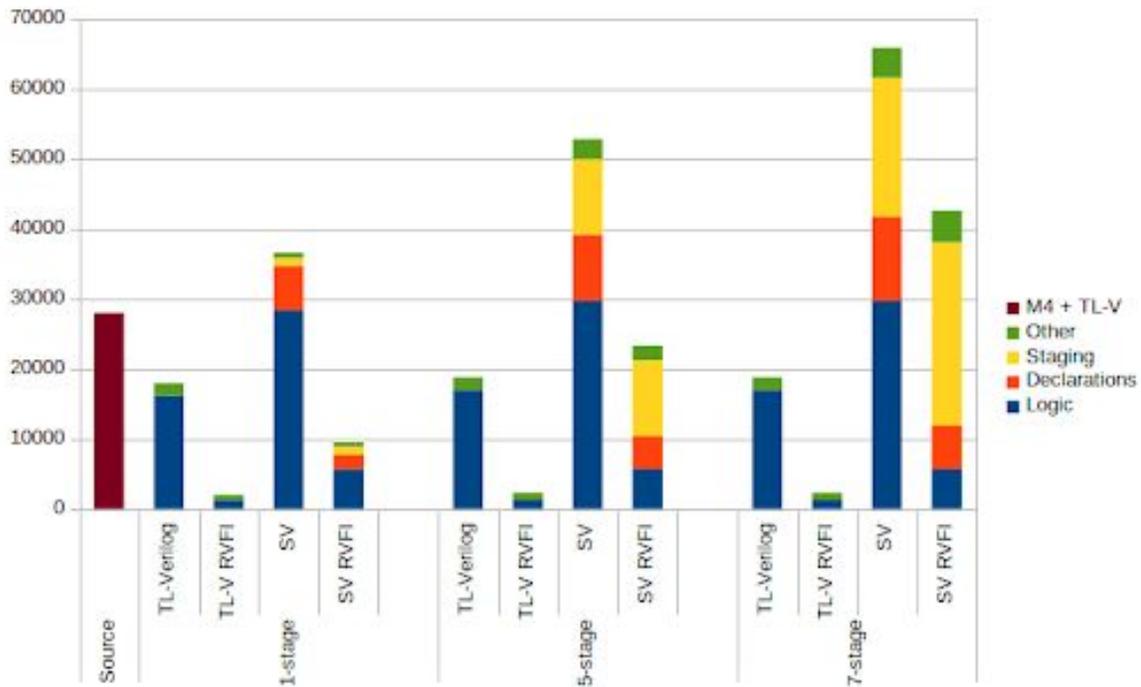

Fig. 4. Code-size comparison among various configurations

To compare the TL-Verilog implementation of WARP-V and its verification harness with a SystemVerilog implementation, Fig. 4 shows character counts of source code for a one-stage, five-stage, and seven-stage implementation of WARP-V. Comments and whitespace are excluded from the data. The WARP-V source code is represented in the leftmost bar. Source code includes parameterization, various selectable components, and code that constructs TL-Verilog code utilizing M4 preprocessing. The TL-Verilog code that results from preprocessing the source code is represented in two separate bars for each implementation--one for the hardware logic and one for the verification harness. The TL-Verilog code generates SystemVerilog code that is also represented in separate bars for hardware and harness.

As suggested by the TL-Verilog hardware logic bars, the TL-Verilog code is almost identical for the various implementations. SystemVerilog hardware code, on the other hand, varies dramatically. The generated staging logic (flip-flops) and associated signal declarations increase with pipeline depth, consistent with examples in [1]. SystemVerilog logic expressions, on the other hand, are comparable across implementations, though, unlike the TL-Verilog code, the SystemVerilog logic expressions, following an industry-recommended coding style, reflect their pipeline stages and therefore differ across implementations.

Similar but more dramatic trends are seen in the harness code. Staging code grows even more dramatically for the harness. In the seven-stage implementation, the test harness approaches the size of the design itself. It is also noteworthy that this SystemVerilog code for just the harness of this single implementation is larger than the entire flexible WARP-V codebase with its harness. It is not uncommon for generated code to experience significant code bloat, however great emphasis has been given to the generation of readability SystemVerilog code conforming to industry-adopted coding style. Code bloat is estimated to be a modest 10 to 15 percent over handwritten code conforming to the same coding style, due primarily to the strict conformance of the tools to the coding conventions. The primary explanation for the dramatic size different is modeling methodology.

It is also meaningful to compare the size of the single TL-Verilog test harness with the combined code that would be required in an RTL methodology for these three implementations (recognizing that any number of implementations are possible). The resulting ratio is 1:70.

## I. Conclusion

A single compact WARP-V codebase generates multiple different and larger SystemVerilog models together with their formal verification harness. The development overhead of a verification harness in an RTL methodology becomes substantial as the pipeline depth increases, and it must be repeated for every implementation. Due to the timing-abstract and transaction-level methodology of TL-Verilog, the TL-Verilog verification harness code is minimal. It is unaffected by the depth of the pipeline and can be utilized across implementations. Flexibility in both the hardware model and the associated verification collateral broadens the applicability of IP and unlocks greater leverage and reuse benefits from SoC design methodology.

## References

[1] S. Hoover, "Timing-Abstract Circuit Design in Transaction-Level Verilog," 2017 IEEE International Conference on Computer Design (ICCD), Boston, MA, 2017, pp. 525-532.

[2] S. Hoover, A. Hadnagy, The WARP-V TL-Verilog RISC-V CPU Core Generator, Available at: https://github.com/stevehoover/warp-v [Last edited 12 Oct. 2018].

[3] Clifford Wolf, et al,, A RISC-V Formal Verification Framework, Available at: https://github.com/cliffordwolf/riscv-formal [Last edited 5 Oct. 2018].

[4] Clifford Wolf, et al., The SymbiYosys Model Checker, Available at: https://github.com/YosysHQ/SymbiYosys [Last edited 12 Sept. 2018].

[5] S. Hoover, et al., "TL-X 2a HDL Extension Draft Syntax Specification, Work in Progress" https://www.TL-X.org, 2018.